\documentclass[aps,prd,twocolumn,eqsecnum,showpacs,nofootinbib]{revtex4}%
\usepackage{amsmath}
\usepackage{subfigure}
\usepackage{float}
\usepackage{graphicx}
\usepackage{subfigure}
\usepackage{footnote}
\usepackage{amsfonts,amssymb,theorem}
\usepackage{bm}
\usepackage{color}
\usepackage{amsfonts}
\usepackage{amssymb}%
\providecommand{\U}[1]{\protect\rule{.1in}{.1in}}
{\theorembodyfont{\upshape}

}
{\theorembodyfont{\upshape}

}
{\theorembodyfont{\upshape}

}
{\theorembodyfont{\upshape}

}
{\theorembodyfont{\upshape}

}
{\theorembodyfont{\upshape}

}
\newcommand{\dalm}{\kern1pt\vbox{\hrule height 0.9pt\hbox{\vrule width
0.9pt\hskip 2.5pt\vbox{\vskip 5.5pt}\hskip 3pt\vrule width 0.3pt}\hrule height
0.3pt}\kern1pt}

\def\b2hat{ {\hat b}_2 }

\begin{document}
\begin{flushright}
UAI-PHY-18/01 \vspace{-1cm}
\end{flushright}
\title{\vskip 9mm Black hole and BTZ-black string in the Einstein-$SU(2)$ Skyrme
model }
\author{Marco Astorino$^{1}$}
\thanks{marco.astorino@gmail.com}
\author{Fabrizio Canfora${}^{2}$}
\thanks{canfora@cecs.cl}
\author{Marcela Lagos$^{3}$}
\thanks{marcelagos@udec.cl}
\author{Aldo Vera$^{3}$}
\thanks{aldovera@udec.cl}
\affiliation{$^{1}$ Departamento de Ciencias, Facultad de Artes Liberales, UAI Physics
Center, Universidad Adolfo Iba\~nez, Av. Padre Hurtado 750, Vi\~na del Mar, Chile}
\affiliation{$^{2}$ Centro de Estudios Cient\'{\i}ficos (CECS), Casilla 1469, Valdivia, Chile}
\affiliation{$^{3}$ Departamento de F\'{\i}sica, Universidad de Concepci\'{o}n, Casilla
160-C, Concepci\'{o}n, Chile}

\begin{abstract}
We present novel analytic hairy black holes with a flat base manifold in the
(3+1)-dimensional Einstein $SU(2)$-Skyrme system with negative cosmological
constant. We also construct (3+1)-dimensional black strings in the Einstein
$SU(2)$-non linear sigma model theory with negative cosmological constant. The
geometry of these black strings is a three-dimensional charged BTZ black hole
times a line, without any warp factor. The thermodynamics of these
configurations (and its dependence on the discrete hairy parameter) is
analyzed in details. A very rich phase diagram emerges.

\end{abstract}

\maketitle

\section{Introduction}

The idea that one can make up fermions out of a purely Bosonic Lagrangian as
solitonic excitations (for a detailed review see \cite{spin}) is one of the
most remarkable results in Quantum Field Theory (QFT henceforth). Skyrme's
theory \cite{skyrme} is the most important example in nuclear and particles
physics. When the Skyrme term is included in the low energy action of Pions,
static soliton solutions with finite energy, called \textit{Skyrmions} (see~
\cite{multis2}-\cite{susy}) describing Fermionic degrees of freedom are
allowed (see \cite{bala2}-\cite{rational} and references therein). The
agreement of the theoretical calculations with experiments is quite good.
However, the Skyrme field equations are very difficult to solve (one reason
being that the Skyrme-BPS bound cannot be saturated in the generic case) and
so, until very recently, basically no analytic solution of the Skyrme field
equations in which one could analyze explicitly the effects of the Skyrme term
was available.

Due to the close relation of the Skyrme theory with the low-energy limit of
QCD, the Einstein-Skyrme system has attracted a lot of attention. The first
important results in this topic were constructed numerically. In particular,
Droz, Heusler, and Straumann \cite{3} (following the findings of Luckock and
Moss \cite{4}) constructed black hole solutions with a non-trivial Skyrme hair
with a spherically symmetric ansatz. Such counterexample to the no-hair
conjecture is also stable against linear perturbations \cite{5}. In \cite{6}
and \cite{7} gravitating solitons and their dynamical features have been also considered.

When the Skyrme coupling constant vanishes, the Skyrme action reduces to the
nonlinear sigma model, which is a very important effective field theory in
itself. The applications of the nonlinear sigma model range from quantum field
theory to statistical mechanics systems, to the quantum hall effect, to super
fluid $^{3}$He and string theory \cite{Mantonbook}. The main use for the
$SU(2)$ nonlinear sigma model is, probably, the description of the low-energy
dynamics of Pions (see for instance \cite{example}, or for a detailed review
\cite{nair}). Therefore, the Einstein nonlinear sigma model system is also a
very important topic, and the construction of analytical solutions is as
relevant as the Einstein-Skyrme system itself, since in the same way, until
recently only numerical solutions had been found\footnote{If a suitable
interaction potential is included, some interesting analytic solutions can be
constructed \cite{anabalon2012}.}.

It is worth to emphasize that the search for analytic solutions in models such
as the Skyrme model, the nonlinear sigma model and their gravitating
counterparts is not just of academic interest. For instance it was a well
known fact, from a numerical point of view, that the Skyrmions in flat spaces
becomes unstable, when a too large isospin chemical potential is introduced.
But only only very recently, in \cite{CG2} and \cite{CG3}, it was derived an
analytic formula for this critical chemical potential which also clarifies the
physical mechanism behind this instability. The hope is that these techniques,
which lead to such important step in the analysis of the Skyrme phase diagram
on flat spaces, will also be useful in clarifying the phase diagrams of hairy
black holes in the Einstein-Skyrme system, as well as in the
Einstein-nonlinear sigma model system. Besides the intrinsic interest of these
phase diagrams, the hairy black holes and black strings solutions which will
be constructed here have potentially many applications in the context of
AdS/CFT correspondence (see \cite{caldavero}, \cite{papanto}\ and references therein).

Using some recent results on the generalization of the hedgehog ansatz to
non-spherically symmetric configurations \footnote{The techniques developed in
these references are very flexible as they apply to the Skyrme model (both
without and with extra moduli degrees of freedom), to the Skyrme-Einstein
system as well as to the Yang-Mills-Higgs theory.} \cite{41}-\cite{ortaggio}\footnote{Similar solutions but in different theories have been found in \cite{Cisterna:2017qrb} and \cite{Cisterna:2017jmv} },
we construct analytic black holes with flat horizons possessing a discrete
hairy parameters: to the best of authors knowledge, these are the first
examples of this type in Einstein-Skyrme theory. The thermodynamics of these
black holes is analyzed in details and a very rich phase diagram is disclosed.

The same techniques also allow to construct black strings in the
(3+1)-dimensional Einstein nonlinear sigma model theory with negative
cosmological constant: the (2+1)-dimensional transversal sections of these
black strings correspond to a charged BTZ black hole. The novel feature of
these black strings is that (unlike what happens, for instance, in the BTZ
black string constructed in \cite{emparan}) the present charged BTZ black
string has no warping factor as the metric is really the direct product of a
charged BTZ with a line.

In the following section we review the Einstein-Skyrme and Einstein nonlinear
sigma model systems. In Sec. III, a pedagogical overview of the generalized
hedgehog ansatz is presented and the matter field and metric ansatz are
constructed. In Sec. IV, the field equations and the solutions for some
interesting cases are shown. In Sec. V, thermodynamics and stability of our
solutions is discussed. Concluding remarks and future prospects are summarized
in the last section. Some useful formulas are collected in the appendix.

\section{The $SU(2)$ Einstein - Skyrme and Einstein - Nonlinear sigma model
systems}

The Skyrme Lagrangian describes the low-energy interactions of pions or
baryons. This observation of Skyrme was, and still is, remarkable because it
provided with the first example of a purely Bosonic Lagrangian able to
describe both bosonic and Fermionic degrees of freedom. The $SU(2)$ Skyrme
field is a $SU(2)$-valued scalar field described by the following action
\begin{equation}
S=S_{\mathrm{G}}+S_{\mathrm{Skyrme}},
\end{equation}
where the gravitational action $S_{\mathrm{G}}$ and the Skyrme action
$S_{\mathrm{Skyrme}}$ are given by
\begin{align}
S_{\mathrm{G}}=  &  \frac{1}{16\pi G}\int d^{4}x\sqrt{-g}(\mathcal{R}%
-2\Lambda),\label{einskyrm}\\
S_{\mathrm{Skyrme}}=  &  \int d^{4}x\sqrt{-g}\mathrm{Tr}\left(  \frac{F_{\pi
}^{2}}{16}R^{\mu}R_{\mu}+\frac{1}{32e^{2}}F_{\mu\nu}F^{\mu\nu}\right)  \ .
\label{sky}%
\end{align}
Here $R_{\mu}$ and $F_{\mu\nu}$ are defined by
\begin{align}
R_{\mu}=  &  U^{-1}\nabla_{\mu}U\ ,\label{skyrme2}\\
F_{\mu\nu}=  &  \left[  R_{\mu},R_{\nu}\right]  ,\ \label{skyrmenotation}%
\end{align}
while $G$ is the Newton constant and the positive parameters $F_{\pi}$ and $e$
are fixed by comparison with experimental data. The Skyrme fields satisfy
physically reasonable reasonable condition, such as the dominant energy
condition~\cite{Gibbons:2003cp}.

For convenience, defining $K=F_{\pi}^{2}/4$ and $\lambda=4/(e^{2}F_{\pi}^{2}%
)$, we write the Skyrme action as
\begin{equation}
S_{\mathrm{Skyrme}}=\frac{K}{2}\int d^{4}x\sqrt{-g}\mathrm{Tr}\left(  \frac
{1}{2}R^{\mu}R_{\mu}+\frac{\lambda}{16}F_{\mu\nu}F^{\mu\nu}\right)  \ .
\label{skyrme-action}%
\end{equation}
The nonlinear sigma model corresponds to the $\lambda\rightarrow0$ limit of
the above action. The resulting Einstein equations are
\begin{equation}
G_{\mu\nu}+\Lambda g_{\mu\nu}=8\pi GT_{\mu\nu}, \label{einskyrmequ}%
\end{equation}
where $G_{\mu\nu}$ is the Einstein tensor and
\begin{align}
T_{\mu\nu}=  &  -\frac{K}{2}\mathrm{Tr}\biggl[\biggl(R_{\mu}R_{\nu}-\frac
{1}{2}g_{\mu\nu}R^{\alpha}R_{\alpha}\biggl)\nonumber\\
&  +\frac{\lambda}{4}\biggl(g^{\alpha\beta}F_{\mu\alpha}F_{\nu\beta}-\frac
{1}{4}g_{\mu\nu}F_{\alpha\beta}F^{\alpha\beta}\biggl)\biggl]\ .
\label{timunu1}%
\end{align}
The Skyrme equations are written as
\begin{equation}
\nabla^{\mu}R_{\mu}+\frac{\lambda}{4}\nabla^{\mu}[R^{\nu},F_{\mu\nu}]=0\ .
\label{nonlinearsigma1}%
\end{equation}
Hence, the full Einstein-sigma model field equations correspond to the
$\lambda\rightarrow0$ limit in Eqs. (\ref{einskyrmequ}), (\ref{timunu1}) and
(\ref{nonlinearsigma1}).

The winding number for a given solution is given by
\begin{equation}
B=\frac{1}{24\pi^{2}}\int\rho_{B}\ ,\quad\rho_{B}=\text{Tr}\left(
\epsilon^{ijk}A_{i}A_{j}A_{k}\right)  \ . \label{winding}%
\end{equation}
It is well known (see, for instance, \cite{Mantonbook}\ and references
therein) that the above integral is a conserved topological charge of the
theory. When the topological density $\rho_{B}$\ is integrated on a space-like
surface, $B$ is the Baryon number of the configuration.

Here $R_{\mu}$ is expressed as
\begin{equation}
R_{\mu}=R_{\mu}^{i} \tau_{i}\ ,
\end{equation}
in the basis of the SU(2) algebra generators
\[
\tau^{k}=i\sigma^{k}\ ,
\]
(where $\sigma^{k}$ are the Pauli matrices, the Latin index $i=1,2,3$
corresponds to the group index, which is raised and lowered with the flat
metric $\delta_{ij}$), which identically satisfy
\begin{equation}
\tau^{i}\tau^{j}=-\delta^{ij}\mathbf{1}-\varepsilon^{ijk}\tau^{k}\ ,
\end{equation}
where $\mathbf{1}$ is the identity $2\times2$ matrix and $\varepsilon_{ijk} $
and $\varepsilon^{ijk}$ are the totally antisymmetric Levi-Civita symbols with
$\varepsilon_{123}=\varepsilon^{123}=1$.

The standard parametrization of the SU(2)-valued scalar $U(x^{\mu})$:%
\begin{equation}
U(x^{\mu})=Y^{0}\mathbf{1}+Y^{i}\tau_{i}\ ,\quad U^{-1}(x^{\mu})=Y^{0}%
\mathbf{1}-Y^{i}\tau_{i}\ , \label{standard1}%
\end{equation}
where $Y^{0}=Y^{0}(x^{\mu})$ and $Y^{i}=Y^{i}(x^{\mu})$ satisfy
\begin{equation}
\left(  Y^{0}\right)  ^{2}+Y^{i}Y_{i}=1\ . \label{standard3}%
\end{equation}
Thus, as expected in the $SU(2)$ case, the theory describes three scalar
degrees of freedom (due to the constraint in Eq. (\ref{standard3})). From the
definition~(\ref{skyrme2}), $R_{\mu}^{k}$ is written as
\begin{equation}
R_{\mu}^{k}=\varepsilon^{ijk}Y_{i}\nabla_{\mu}Y_{j}+Y^{0}\nabla_{\mu}%
Y^{k}-Y^{k}\nabla_{\mu}Y^{0}\ . \label{standard4}%
\end{equation}

Another convenient way (which will be used in the following) to describe
$SU(2)$-valued scalar field uses the Euler angle representation (for a
detailed review see \cite{euler}). In this representation, the most general
$SU(2)$-valued scalar field can be written as%
\begin{equation}
U(x^{\mu})=e^{\tau_{3}u_{1}(x^{\mu})}e^{\tau_{2}u_{2}(x^{\mu})}e^{\tau
_{3}u_{3}(x^{\mu})}\ . \label{eulerepr}%
\end{equation}
As it happens in the standard representation for $SU(2)$-valued scalar field
(in Eqs. (\ref{standard1}) and (\ref{standard3})), in the Euler angle
representation\footnote{If necessary, one can pass from one representation to
the other (as it is a standard computation to express $u_{1}(x^{\mu})$,
$u_{2}(x^{\mu})$ and $u_{3}(x^{\mu})$ in terms of the $Y_{0}$ and $Y_{i}$ in
Eqs. (\ref{standard1}) and (\ref{standard3}) using, for instance, the results
in \cite{euler}). However, the novel results presented here are more easily
expressed in the Euler angle representation.} in Eq. (\ref{eulerepr}) there
are three scalar degrees of freedom: the three scalar functions $u_{1}(x^{\mu
})$, $u_{2}(x^{\mu})$ and $u_{3}(x^{\mu})$. Thus, in order to solve the Skyrme
field equations in the Euler angle representation one needs to construct a
good ansatz for $u_{1}(x^{\mu})$, $u_{2}(x^{\mu})$ and $u_{3}(x^{\mu})$: we
will outline the strategy to build such an ansatz in the next section.

\section{Matter field and metric ansatz}

A very important class of black holes both from the strictly theoretical
viewpoint as well as from the point of view of holographic applications
corresponds to hairy black holes with flat horizons and negative cosmological
constant (see \cite{caldavero}, \cite{papanto}\ and references therein). The
interest in black holes with hairy parameters arises from the fact that often
such black holes exhibit a very complex thermodynamical behavior. The interest
in having flat horizons with negative cosmological constant lies in the
possibility to describe, via the AdS/CFT correspondence, very interesting
field theories on the boundary of the black hole space-time itself. It is
usually quite difficult to construct hairy black holes in sectors of the
standard model minimally coupled with General Relativity. These considerations
are behind our interest in constructing this type of configurations within the
Einstein-Skyrme system as it describes the minimal coupling of (the low energy
limit of) QCD with General Relativity.

\subsection{The main theoretical tool: the generalized hedgehog ansatz}

In this subsection, the concept of hedgehog ansatz in the Einstein-Skyrme
system will be shortly described. The technical difficulty to construct
analytic black hole configurations in the Einstein-Skyrme system arises from
the fact that already the Skyrme field equations on flat space-times in
themselves are a very difficult nut to crack (see \cite{manton}\ and
references therein). Thus, one may argue that the situation in the coupled
Einstein-Skyrme system is even worse. In fact, quite recently an effective
strategy suitable to deal with this type of problems has been developed in the
references \cite{41}-\cite{ortaggio}. Such a strategy is divided into two steps.

\textit{The first step}: identify the symmetries of the space-times of
interest in such a way to distinguish clearly the Killing coordinates from the
non-Killing coordinates of the metric.

\textit{The second step}: choose the $SU(2)$ valued ansatz $U$ in such a way
that it \textit{also depends} on the Killing coordinates of the
metric\footnote{It is worth to emphasize that, in the simple case of one
scalar field (without internal symmetries) minimally coupled with general
relativity, this is not what one would do. Consider, for instance, a static
spherically symmetric space-time. The most obvious ansatz for the scalar field
would be to assume that the scalar field only depends on the (non-Killing)
radial coordinate.} of interest with the additional (very important) condition%
\begin{equation}%
\mathcal{L}%
_{\overrightarrow{K}}U\neq0\ ,\ \
\mathcal{L}%
_{\overrightarrow{K}}T_{\mu\nu}^{U}=0\ , \label{generhedgedef}%
\end{equation}
where $%
\mathcal{L}%
_{\overrightarrow{K}}$\ is the Lie derivative along the Killing fields
(denoted by $\overrightarrow{K}$)\ of the metric while $T_{\mu\nu}^{U}$\ is
the energy-momentum tensor (defined in Eq. (\ref{timunu1})) corresponding to
the $SU(2)$ valued ansatz $U$ itself. The possibility to implement the above
strategy arises from the non-trivial internal symmetry group of the field
theory minimally coupled with gravity.

The requirement in Eq. (\ref{generhedgedef}) asks to find an ansatz
\textit{which is not invariant under the symmetries of the metric} but which,
nevertheless, possesses an energy-momentum tensor which is compatible with the
symmetries of the space-time of interest. Such a condition is somehow rigid
since often it allows to determine the functional form of the ansatz itself
almost completely.

Once Eq. (\ref{generhedgedef}) has been satisfied, it is usually a quite easy
task to verify whether or not there is still enough freedom left in $U$ to be
able to solve (at least numerically) the Skyrme field equations in the metric
of interest.

Thus, the above two steps (and in particular Eq. (\ref{generhedgedef}))
summarize the \textit{generalized hedgehog ansatz}.

The word "generalized" arises from the following fact. In the original papers
by Skyrme \cite{skyrme}, the spherically symmetric hedgehog ansatz was%
\begin{align}
U_{S}(x^{\mu})  &  =\cos\left(  \alpha(r)\right)  \mathbf{1}+\sin\left(
\alpha(r)\right)  n^{j}\tau_{j}\ ,\label{ori1}\\
ds^{2}  &  =-dt^{2}+dr^{2}+r^{2}\left(  d\theta^{2}+\sin^{2}\theta
d\varphi^{2}\right)  \ ,\label{ori2}\\
n^{1}  &  =\sin\theta\cos\varphi\ ,\ n^{2}=\sin\theta\sin\varphi
\ ,\ n^{3}=\cos\theta\ , \label{ori3}%
\end{align}
where $\alpha(r)$ is the so-called Skyrmion profile. Although Skyrme arrived
at his ansatz following a different reasoning, it is a direct computation to
verify that the ansatz in Eqs. (\ref{ori1}) and (\ref{ori3}) satisfies Eq.
(\ref{generhedgedef}) in which the Killing fields $\overrightarrow{K}$
correspond to the $SO(3)$ rotations of the flat metric in Eq. (\ref{ori2}). In
other words, the Skyrme ansatz could have been found solving Eq.
(\ref{generhedgedef}) in a spherically symmetric metric in which the Killing
fields $\overrightarrow{K}$ correspond to the $SO(3)$ rotations. Once the
functional form of the ansatz has been restricted by Eq. (\ref{generhedgedef}%
), one can plug in it into the three Skyrme field equations Eq.
(\ref{nonlinearsigma1}) corresponding to the metric with the Killing fields
$\overrightarrow{K}$. In the original case analyzed by Skyrme, it can be
directly verified that when one plugs Eqs. (\ref{ori1}) and (\ref{ori3}) into
Eq. (\ref{nonlinearsigma1}) (for the metric in Eq. (\ref{ori2})) the three
Skyrme field equations become proportional, so that the full system of three
coupled field equations reduces to just one scalar equation for the Skyrmion
profile $\alpha(r)$.

All the above very convenient properties of the original spherical hedgehog
ansatz are well known of course. However, what was not widely appreciated in
the literature is that one can construct ansatz with similar nice properties
even without spherical symmetry. The key point is that the condition in Eq.
(\ref{generhedgedef}) makes sense in more general situations than spherically
symmetric space-time. Indeed, this simple observation in \cite{41} \cite{46}
allowed to find the first non-trivial analytic solutions in Skyrme and
Einstein-Skyrme theories in \cite{56} \cite{58} \cite{ACZ} \cite{CG2}
\cite{CG3}. These are the reasons behind the name \textit{generalized hedgehog
ansatz}.

In the present case, there is an additional technical problem. We are
interested in hairy black holes, thus we look for configurations possessing
neither topological nor Noether charges related with the isospin symmetry.
This issue will be analyzed in the next subsection.

\subsubsection{An example}

Before going into the details of the novel results, here we will describe an
example (which corresponds to the first analytic gravitating Skyrmions in
(3+1)-dimensional Einstein-Skyrme system found in \cite{ACZ}) in which the
strategy outlined above works perfectly. Let us consider the following
space-time metric (the first step of the strategy)
\begin{equation}
ds^{2}=-dt^{2}+\rho\left(  t\right)  ^{2}\left[  (d\gamma+\cos\theta
d\varphi)^{2}+d\theta^{2}+\sin^{2}\theta d\varphi^{2}\right]  \ ,
\label{simplem}%
\end{equation}%
\begin{equation}
0\leq\gamma<4\pi\ ,\;\;\;\ 0\leq\theta<\pi\ ,\;\;\;\ 0\leq\varphi<2\pi\ .
\label{range1}%
\end{equation}
The spatial $\left(  t=const\right)  $ sections of the above metric are
three-spheres. Consequently, the above metric possesses all the Killing fields
of the three-sphere (and $\gamma$, $\theta$\ and $\varphi$\ can be considered
to be Killing coordinates). The only non-trivial "non-Killing" coordinate is
the time $t$.

The second step of the strategy corresponds to find an ansatz of $U\in SU(2)$
such that%
\[%
\mathcal{L}%
_{\overrightarrow{K}}U\neq0\ ,\ \
\mathcal{L}%
_{\overrightarrow{K}}T_{\mu\nu}^{U}=0\ ,
\]
where $\overrightarrow{K}$\ are the Killing field of the three-sphere. The
solution to the above condition is given by%
\[
U(x^{\mu})=Y^{0}(x^{\mu})I\pm Y^{i}(x^{\mu})t_{i},\qquad\left(  Y^{0}\right)
^{2}+Y^{i}Y_{i}=1\ \ ,
\]
\begin{align}
\label{sessea1} &  Y^{0}=\cos{\alpha}\ ,\quad Y^{i}=n^{i}\sin{\alpha}\ ,\\
&  n^{1}=\sin{\Theta}\cos{\Phi}\ ,\quad n^{2}=\sin{\Theta}\sin{\Phi}\ ,\quad
n^{3}=\cos{\Theta}\ \nonumber
\end{align}
where
\begin{equation}
\Phi=\frac{\gamma+\varphi}{2}\ ,\ \tan\Theta=\frac{\cot\left(  \frac{\theta
}{2}\right)  }{\cos\left(  \frac{\gamma-\varphi}{2}\right)  }\ ,\ \tan
\alpha=\frac{\sqrt{1+\tan^{2}\Theta}}{\tan\left(  \frac{\gamma-\varphi}%
{2}\right)  }\ . \label{sessea2}%
\end{equation}
As it has been already emphasized, at a first glance the situation is quite
dangerous since the condition in Eq. (\ref{generhedgedef}) is rather rigid as,
in this example, it fixes the ansatz for the $SU(2)$-valued scalar field
completely while the Skyrme field equations have not been considered yet!
Nevertheless, remarkably (as it was shown in \cite{ACZ}), when one plugs the
ansatz in Eqs. (\ref{sessea1}) and (\ref{sessea2}) into the Skyrme field
equations in Eq. (\ref{nonlinearsigma1}) in the metric in Eq. (\ref{simplem}),
\textit{the Skyrme field equations are identically satisfied}. Thus, despite
the fact that the condition in Eq. (\ref{generhedgedef}) gives rise to an
ansatz in which, basically, there is no freedom left, the resulting ansatz is
very well suited to solve the Skyrme field equations. Thus, we are only left
with the problem to solve the Einstein equation with the energy-momentum
tensor in Eq. (\ref{timunu1}) corresponding to the ansatz in Eqs.
(\ref{sessea1}) and (\ref{sessea2}). In fact, the generalized hedgehog
strategy has been designed in such a way that this last step is compatible:
the condition $%
\mathcal{L}%
_{\overrightarrow{K}}T_{\mu\nu}^{U}=0$ precisely ensures that the resulting
energy-momentum tensor is a consistent source for the metric in Eq.
(\ref{simplem}). A direct computation shows \cite{ACZ} that the
Einstein-Skyrme equations reduce\footnote{Note that these equations
corresponds to the ones in reference \cite{ACZ} by rescaling $\rho
\rightarrow2\rho$.} in this example to
\begin{equation}
\rho^{\prime2}=\frac{\Lambda}{3}\rho^{2}+\frac{\lambda\kappa K}{32\rho^{2}%
}+\frac{\kappa K-2}{8}\ ,\quad\rho^{\prime\prime}=\frac{\Lambda}{3}\rho
^{2}-\frac{\lambda\kappa K}{32\rho^{3}}\ , \label{isoskSU(2)}%
\end{equation}
where $(^{\prime})$ denotes derivative with respect to the time coordinate,
$t$. In the following sections, it will be shown that this strategy works very
well even when the metric of interest has different symmetries.

\subsection{The concrete ansatz for the novel solutions}

The first step of the generalized hedgehog strategy is to identify the
symmetries of the class of metric of interest. In the present case, the
natural metric ansatz describing both black holes with flat horizons and black
strings is
\begin{equation}
ds^{2}=-A(r)dt^{2}+B(r)dr^{2}+C(r)d\theta^{2}+D(r)d\phi^{2}\ ,
\label{metricflat1}%
\end{equation}
where the range of the angular coordinates can be fixed as
\begin{equation}
0\leq\theta\leq\pi\ ,\ 0\leq\phi\leq2\pi~. \label{range}%
\end{equation}

The second step requires to solve Eq. (\ref{generhedgedef}) for a $SU(2)$
valued scalar field which also depend on the Killing coordinates $\theta$ and
$\phi$ of the metric in Eq. (\ref{metricflat1}). Form the viewpoint of the
generalized hedgehog approach, the simplest possibility is actually to search
for an ansatz $U$ which\textit{ only} depends on such Killing coordinates
$\theta$ and $\phi$. As it will be now shown, this approach does lead to novel
and interesting solutions. A further motivation behind this choice is that we
are interested in black holes with hairy parameters. These configuration are
easier to identify in case of absence of extra parameters related to
topological or Noether charges coming from the $SU(2)$ symmetry of the matter
field. The simplest way to avoid the presence of a non-vanishing topological
charge is to allow the matter field to only depend on two coordinates (which,
according to the generalized hedgehog strategy, should then be Killing
coordinates) as in this case $\rho_{B}$ in Eq. (\ref{winding}) vanishes identically.

In the cases in which the Killing fields of the metric of interest are
commuting (as for the metric in Eq. (\ref{metricflat1})) it is more convenient
to use the Euler angle representation in Eq. (\ref{eulerepr}). The simplest
non-trivial possibility is
\begin{equation}
U=e^{\tau_{2}u_{2}(\theta,\phi)}e^{\tau_{3}u_{3}(\theta,\phi)}\ , \label{ans1}%
\end{equation}
with $u_{i}$ two real functions of the Killing coordinates $\theta$ and $\phi$
so that, obviously, one has%
\[%
\mathcal{L}%
_{\overrightarrow{K}}U\neq0
\]
where
\[
\overrightarrow{K}=\left(  \partial_{\theta}\ ,\ \partial_{\phi}\right)  \ .
\]
It is worth to emphasize here that the ansatz in Eq. (\ref{ans1}) does not
possess Noether charges associated to the internal symmetry group of the
theory (the global Isospin group $SU(2)$ in the present case). The reason is
that such a Noether charge would be the spatial integral of the time-component
of the corresponding Noether current. On the other hand, the time-component of
the Noether current is proportional to%
\begin{align*}
J_{t}  &  \sim U^{-1}\partial_{t}U+\frac{\lambda}{4}[U^{-1}\nabla^{\nu
}U,F_{t\nu}]\ ,\\
F_{t\nu}  &  =[U^{-1}\partial_{t}U,U^{-1}\partial_{\nu}U]\ ,
\end{align*}
so that it vanishes identically since the configuration is static. Hence, the
ansatz for the scalar field we will consider here does carry neither
topological nor Noether charges.

The (second part of the) second step of the strategy is now to solve the
following equation%
\[%
\mathcal{L}%
_{\overrightarrow{K}}T_{\mu\nu}^{U}=0
\]
in which $T_{\mu\nu}^{U}$\ is the energy-momentum tensor (defined in Eq.
(\ref{timunu1})) corresponding to the ansatz in Eq. (\ref{ans1}) and to the
metric in Eq. (\ref{metricflat1}). The solution to the above condition is
given by $u_{2}(\theta)=b_{1}\theta/2$ and $u_{3}(\phi)=b_{2}\phi/2$. In
matrix form, this solution corresponds to the following $U$:
\begin{equation}
U=\left(
\begin{array}
[c]{cc}%
e^{\frac{ib_{2}\phi}{2}}\cos(\frac{b_{1}\theta}{2}) & e^{-\frac{ib_{2}\phi}%
{2}}\sin(\frac{b_{1}\theta}{2})\\
-e^{\frac{ib_{2}\phi}{2}}\sin(\frac{b_{1}\theta}{2}) & e^{-\frac{ib_{2}\phi
}{2}}\cos(\frac{b_{1}\theta}{2})
\end{array}
\right)  \ ,\ b_{i}\in\
\mathbb{R}
\ . \label{ans2}%
\end{equation}
Once again, as it also occurred in the example described in the previous
subsection, the ansatz produced by the present strategy is quite rigid and, in
this case, only two integration constants ($b_{1}$ and $b_{2}$) are left.

Nevertheless, also in this case a direct computation reveals that \textit{the
Skyrme field equations Eq. }(\ref{nonlinearsigma1})\textit{\ with the ansatz
in Eqs. }(\ref{ans1})\textit{\ and} (\ref{ans2})\textit{\ are identically
satisfied in any metric of the form in Eq. }(\ref{metricflat1}). This is the
big technical achievement of the generalized hedgehog strategy developed in
\cite{41}-\cite{ortaggio}: it allows to reduce the full Einstein-Skyrme system
just to the Einstein equations with the energy-momentum tensor of the Skyrme
field (as the Skyrme field equations, which usually are the difficult part of
the problem, are identically satisfied). Moreover, by construction, the
energy-momentum tensor is compatible with the symmetries of the metric of
interest (due to Eq. (\ref{generhedgedef})).

\subsubsection{The topology of the horizon}

In this article we will be mainly interested in compact horizons, therefore we
need to impose (anti)periodic boundary conditions for the Skyrme field (see
\cite{manton}, \cite{Marmo}), that is, any solution of the form in Eq.
(\ref{ans2}) of the Skyrme field equations in the metric Eq.
(\ref{metricflat1}) must satisfy
\[
U(\theta,\phi)=\pm U(\theta+n\pi,\phi+2m\pi)\ ,
\]
with $n$, $m$ integers. In order to have a well defined $U$ the integration
constants $b_{i}$ must be integers numbers\footnote{In principle, one can
choose a different range for the coordinates to rescale these values with the
metric functions and coordinate definitions. Nevertheless, being integrating
constant of the matter field, the $b_{i}$ parameters cannot be completely
reabsorbed from the solution. We leave them explicitly to better understand
their physical role and relevance, as we see below.},
\begin{equation}
\{b_{1}\ ,\ b_{2}\}\ \in\mathbb{N}\ . \label{quant1}%
\end{equation}
These $b_{i}$ parameters, as can be seen from (\ref{ans2}), are related to
number of coverings of the $SU(2)$ group.\newline The angles identifications
we are considering determines not only the topology of the event horizon but
also the global topology of the spacetime, even in the asymptotic region. This
means that even though the metric, for large values of the radial coordinate,
behaves locally as the Anti-de-Sitter space, as can be easily seen from the
curvature tensors, the asymptotic region does not recover globally the full
$AdS_{4}$ spacetime. Therefore when the radial coordinate goes to infinity the
spacetime we are studying here are only asymptotically locally
Anti-de-Sitter.\newline The great physical interest of these configurations is
that they show very clearly that the Skyrme contribution to the action is
quite relevant even for purely Pionic configurations without Baryon charge.
For instance, as it will be shown below, it gives rise to a black hole metric
with a $1/r^{2}$-term which mimics the presence of a Maxwell source.

\section{Analytical solutions}

In the previous section we have reduced consistently the full Einstein-Skyrme
system for the metric in Eq. (\ref{metricflat1}) and the Skyrme ansatz in Eqs.
(\ref{ans1}) and (\ref{ans2}) to the Einstein equations with the Skyrme
energy-momentum tensor corresponding to the ansatz in Eqs. (\ref{ans1}) and
(\ref{ans2}). In principle there are four coupled nonlinear differential
equations (see Appendix A), but it is possible to show that one of these is a
combination of the others because of the Bianchi's identity and the form of
the metric ansatz for the matter field. In this section we show some relevant
cases where it is possible to integrate the system analytically. In what
follows we will consider $B(r)=1/A(r)$ and $\kappa= 8\pi G$.

\subsection{Hairy Black hole}

\begin{figure}[h]
\centering
\includegraphics[scale=0.32]{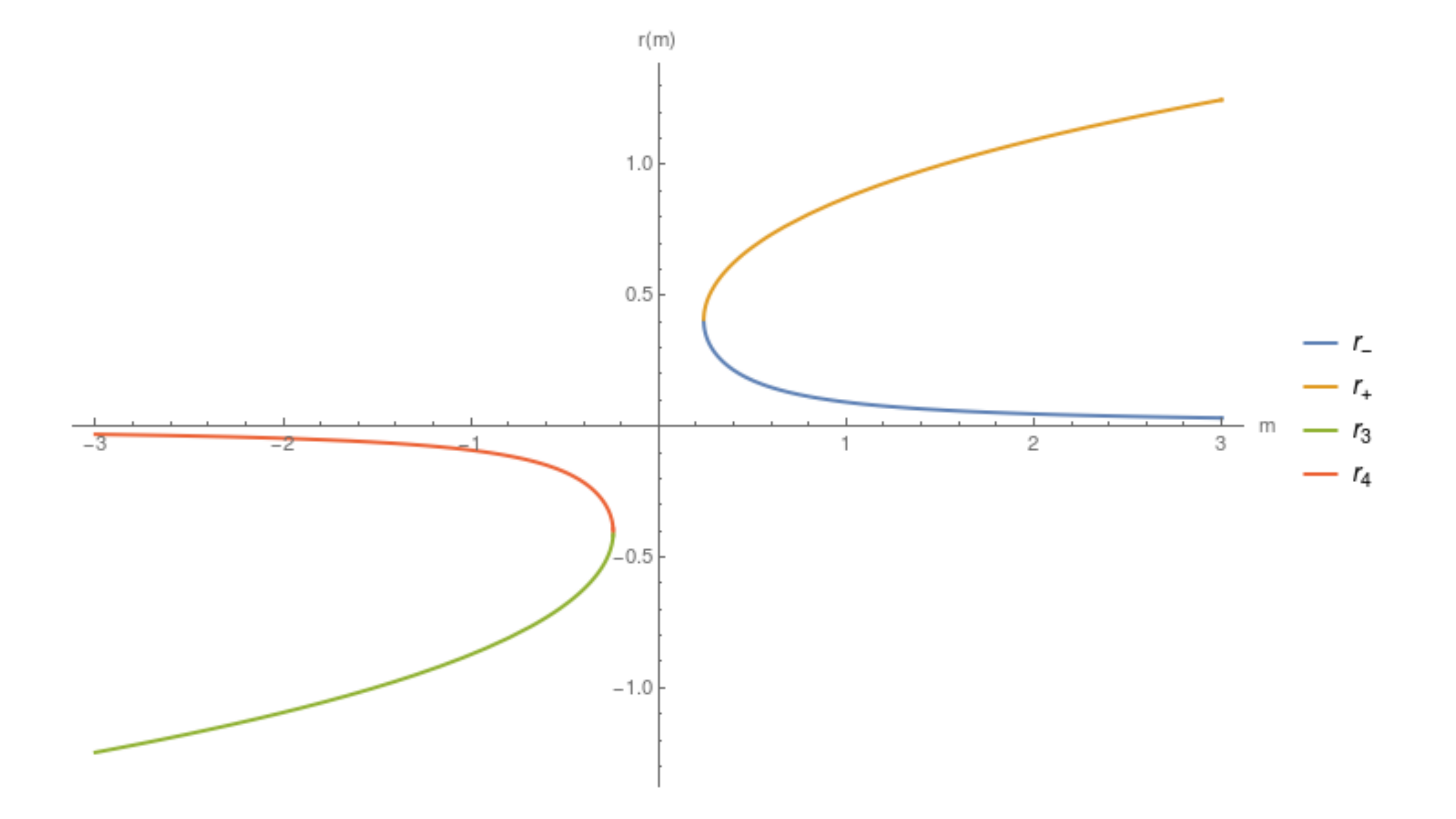}\caption{{\protect\small The event
horizon $r_{+}$ as function of the mass parameter $m$ is portrayed in the
yellow line, while the inner horizon $r_{-}$ is drawn in blue. The value of
the mass, where the blue and yellow line touches, represent the extremal case.
The other two roots $r_{3},r_{4}$ give negative radial distance, hence, are
not physically relevant. The numerical values of the coupling constants and of
the physical parameters of the solution, for the above image, were chosen as
follows $b_{1}=b_{2}=1,\lambda=3,K=1,\kappa=1,\Lambda=-5$. }}%
\label{roots-bh-fig}%
\end{figure}

If we take $C(r)=r^{2}$ and $D(r)=\frac{b_{2}^{2}}{b_{1}^{2}}r^{2}$, the field
equations leads to
\[
A(r)=-\frac{b_{1}^{2}\kappa K}{4}-\frac{m}{r}-\frac{\Lambda}{3}r^{2}%
+\frac{b_{1}^{4}\kappa K\lambda}{32}\frac{1}{r^{2}}\ ,
\]
where $m$ is an integration constant. The metric
\begin{align}
\label{metric-bh}ds^{2}  &  =-\left(  -\frac{b_{1}^{2}\kappa K}{4}-\frac{m}%
{r}-\frac{\Lambda}{3}r^{2}+\frac{ b_{1}^{4}\kappa K\lambda}{32 r^{2} }\right)
dt^{2}\\
&  +\frac{dr^{2}}{-\frac{b_{1}^{2}\kappa K}{4}-\frac{m}{r}-\frac{\Lambda}%
{3}r^{2}+\frac{ b_{1}^{4}\kappa K\lambda}{32 r^{2} }} +r^{2}d\theta^{2}+
\frac{b_{2}^{2}}{b_{1}^{2}} r^{2} d\phi^{2},\nonumber
\end{align}
represents a hairy black hole with flat horizon. This solution reduce to the
black hole of \cite{ortaggio} when $\lambda=0$, while it is the natural flat
horizon generalization of the spherical metric found in \cite{46}. Black hole
with flat horizons are especially relevant in view of their holographic
applications (see, for instance, \cite{emparan}). For $\lambda=0$ there is
only one real root for the $A(r)$ function which corresponds to the event
horizon $r_{+}$
\begin{equation}
\label{rpl0}r_{+} = \frac{b_{1}^{2} \kappa K \Lambda- \left(  12 m\Lambda^{2}
+ \sqrt{\Lambda^{3} (b_{1}^{6}(\kappa K)^{3} +144 m^{2} \Lambda)} \right)
^{2/3}}{2 \Lambda\left(  12 m\Lambda^{2} + \sqrt{\Lambda^{3} (b_{1}^{6}
(\kappa K)^{3} +144 m^{2} \Lambda)} \right)  ^{1/3}} \ .
\end{equation}
For $\lambda\neq0$, the roots of $A(r)$ can also be found analytically, but
they are more involved than the ones of (\ref{rpl0}), because the algebraic
equation becomes of fourth order, thus is more instructive to draw them as
function of the mass parameter $m$ as in Figure \ref{roots-bh-fig}.
Interestingly enough, as is shown in the figure, neither the mass
of the black hole nor the event horizon radius can be arbitrarily small
because the event horizon is not defined for small masses, unlikely the
standard general relativity case. \newline Even in the $\lambda=0$ case, where
there is only one killing horizon, from Eq. (\ref{rpl0}) one infers that the
mass parameter should satisfy $m \geq\frac{b_{1}^{3} (\kappa K)^{3/2}}%
{12\sqrt{-\Lambda}} $ in order the square root to be real. Therefore, the
event horizon cannot be arbitrarily small either but, for the extremal value
of the mass parameter, which saturate the previous inequality, it can be
reduce at most to $\bar{r}_{+}= \frac{b_{1} \sqrt{\kappa K}}{\sqrt{-\Lambda}}$.

This latter feature of the $b_{i}$ parameter resemble the electric charge of
the Reissner-Nordstrom solution, but, in the Skyrme case, the parameters
$b_{i}$ must be quantized due to the boundary conditions satisfied by the
Skyrme field. In particular, $b_{1}$ plays the role of a discrete hairy
parameter (since, as we observed in the previous section, the Skyrme
configurations in Eq. (\ref{ans2}) possess neither Noether charges nor
topological charges). According to this picture the hair cannot be considered
neither of a primary type, because it cannot variate continuously, nor
secondary type because it is not completely fixed. We might consider it
belonging to an intermediate class; a sort of \textit{semi-primary} hair.
\newline The non-removability of $b_{1}$ resembles, to some extent, the role
played by the mass parameter of the three-dimensional BTZ black hole
\cite{btz}: in that case, once the azimuthal coordinate is fixed, it cannot be
reabsorbed by a coordinate transformation to get global AdS space-time.
Moreover, in the next section, we will show how the thermodynamics potentials
depends crucially on the hair parameters. \newline Depending on the
identifications of the coordinates ($\theta,\phi$), the base manifold can be
considered open or compact. When the extremal points of the range of
($\theta,\phi$) are identified, the base manifold becomes a topological torus
$\mathcal{S}^{1}\times\mathcal{S}^{1}$ with area $\mathcal{A}=2\pi^{2}%
r_{+}^{2}\frac{b_{2}}{b_{1}}$. In that case the ratio $b_{1}/b_{2}$ determines
the geometry of the toroidal base manifold, being $b_{1}/b_{2}$ its
Teichmuller parameter. Therefore, the discrete parameters $b_{1}$ and $b_{2}$
can be reabsorbed by rescaling properly the coordinates and the integration
constants only at the price of deforming the geometry of the base manifold and
the Skyrmionic field.\newline The spacial infinity region is asymptotically
locally AdS.\newline When the Skyrmionic parameters coincide, $b_{2}=b_{1}$,
one can take the limit $b_{1}\rightarrow0$. In that case the contribution of
the matter field vanishes and the pure gravitational black hole solution of
\cite{lemos} is recovered.

\subsection{Charged-like BTZ-black string}

When we choose $C(r)=r^{2}$ and $D(r)=L^{2}$ (with $L$ an arbitrary constant),
the Einstein equations are satisfied only in the sector $\lambda=0$, and leads
to
\begin{align*}
A(r)=-\mu-\frac{b_{1}^{2}\kappa K}{4}\log(r)-\frac{\Lambda}{2}r^{2}\ ,  &
\quad L^{2}=-\frac{b_{2}^{2}\kappa K}{4\Lambda}\ .
\end{align*}
In the case in which the cosmological constant is negative and the integration
constant $\mu>0$, the resulting metric reads
\begin{align}
\label{metric-btz}ds^{2}  &  =-\left(  -\mu-\frac{b_{1}^{2}\kappa K}{4}%
\log(r)+\frac{\left\vert \Lambda\right\vert }{2}r^{2}\right)  dt^{2}\\
&  +\frac{1}{-\mu-\frac{b_{1}^{2}\kappa K}{4}\log(r)+\frac{\left\vert
\Lambda\right\vert }{2}r^{2}}dr^{2}+r^{2}d\theta^{2}+\frac{b_{2}^{2}\kappa
K}{4\left\vert \Lambda\right\vert }d\phi^{2}\ .\nonumber
\end{align}
This solution corresponds to a black string with one compactified direction
(namely $\phi$) whose compactification radius $L=\frac{b_{2} \sqrt{\kappa K}%
}{2 \sqrt{|\Lambda|}}$ has been fixed by the field equations. The
three-dimensional metric (corresponding to the $\phi=const$\ hypersurfaces)
resemble the charged BTZ black holes \cite{Martinez:1999qi} with mass $\mu$
and square charge $b_{1}^{2}\kappa K/4$. \newline The nonlinear sigma model
induce an effective electric charge in the three-dimensional metric defining
the black string. It is worth to note that, unlike what happens for instance
in the BTZ black string constructed in \cite{emparan}, the present charged BTZ
black string has no warping factor, as the metric is really the direct product
of a charged BTZ with a one-dimensional line (which can be also considered
compatified in a $S^{1}$ circle). It is also worth to note that it is not
possible to turn off the nonlinear sigma model, to obtain a pure gravitational
solution, as the $S^{1}$ factor would be singular. Thus, the parameter $b_{1}$
plays the role of an effective electric charge while the parameter $b_{2}$
determines the size of the compactified direction of the black string.\newline
As the great majority of charged black holes, this string posses, in general,
both a inner and outer horizon $R_{\pm}$. Unfortunately, due to the presence
of the transcendental function in $A(r)$, the position of the horizon cannot
be written with elementary functions, but only through the
Lambert-$\mathcal{W}$ function (also known as the ProductLog function) in this
way
\begin{align}
R_{+}  &  = \frac{b_{1} \sqrt{\kappa K}}{2} \sqrt{\frac{1}{\Lambda}
\ \mathcal{W}_{-1} \left[  \frac{4 \Lambda}{b_{1}^{2} \kappa K} \exp\left(
\frac{-8 \mu}{b_{1}^{2} \kappa K} \right)  \right]  } \ ,\\
R_{-}  &  = \frac{b_{1} \sqrt{\kappa K}}{2} \sqrt{\frac{1}{\Lambda}
\ \mathcal{W}_{0} \left[  \frac{4 \Lambda}{b_{1}^{2} \kappa K} \exp\left(
\frac{-8 \mu}{b_{1}^{2} \kappa K} \right)  \right]  } \ \ .
\end{align}

\begin{figure}[h]
\includegraphics[scale=0.63]{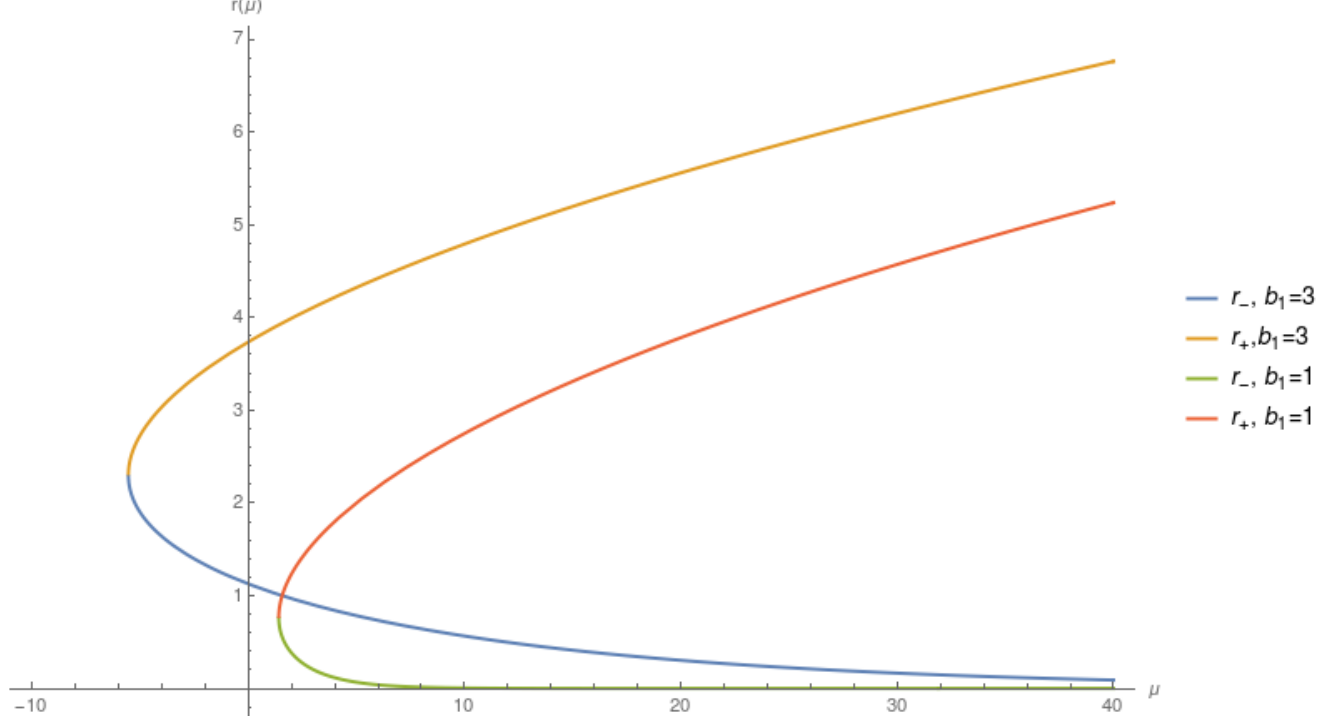}\caption{{\protect\small Killing
horizons of the black string, as function of the mass parameter $\mu$, are
pictured for $b_{1}=1$ and $b_{1}=3$. In both cases the event horizon $r_{+}$
has a lower bound, for certains values of $b_{1}$ the extremal case is not
physically accessible for positive values of the mass parameter $\mu$. }}%
\label{roots-string-fig}%
\end{figure}

As can be seen from the Figure \ref{roots-string-fig}, as in the previous
subsection black hole case, the event horizon $R_{+}$ cannot vanish, as it
happens in the pure gravity case. In the presence of the Skyrmionic matter
$R_{+}$, depending on the values of the parameter $b_{1}$, can only reach a
positive minimum value. 

\section{Thermodynamics}

A very interesting topic is the proper dynamical stability analysis of the
present black holes and black strings solutions. The main difficulty is
revealed by a direct computation of the fully coupled linearized
Einstein-Skyrme field equations in the black hole and black string background
solutions. When the Skyrme field equations are taken into account (due to
their matrix-valued and non-linear nature), the linearized field equations
cannot be reduced to a single master Schrodinger-like equation for the
perturbations (unlike what happens in many situations without matter fields).
This fact prevents any analytic attempt to analyze dynamical stability. Thus,
one has to solve numerically the matrix-valued linearized field equations
around the black hole and black string background solutions. However, this
point is very difficult even from the numerical point of view and it requires
suitable generalizations of the methods available in the literature. We hope
to come back on this issue in a future publication.

On the other hand, as it is well known, the analysis of thermodynamics of
black holes and black strings solutions (besides to be very interesting in
itself) provides with very good qualitative indications on the possible
appearance of instabilities. Thus, in this section we study the thermodynamic
of these solutions and also perform a thermodynamical stability analysis
through comparison between the thermodynamic potentials.

\subsection{Mass, temperature and entropy}

The Hawking temperature, related to the surface gravity $\kappa_{s}$ , is
given by
\begin{equation}
T=\frac{\kappa_{s}}{2\pi}=\frac{1}{2\pi}\sqrt{-\frac{1}{2}\nabla_{\mu}%
\chi_{\nu}\nabla^{\mu}\chi^{\nu}}\ \Bigg|_{r_{+}}=\ \frac{A^{\prime}(r_{+}%
)}{4\pi}\ ,\
\end{equation}
where $\chi$ is the timelike killing vector $\partial_{t}$. From the
Bekenstein-Hawking formula, we can take the entropy as a quarter of the area
\begin{equation}
S=\frac{\mathcal{A}}{4G}\ .\
\end{equation}
To compute the mass will use the standard ADM result, as in \cite{vanzo}.
Consider $\mathcal{S}$ as the two-dimensional spacelike surface at spatial
radial infinity at constant time. The radial orthonormal vector to
$\mathcal{S}$ is given by $n_{\mu}=\left(  0,\sqrt{g_{rr}},0,0\right)  $, the
extrinsic curvature of $\mathcal{S}$ is given by $\mathcal{K}_{\mu\nu}%
=\nabla_{\mu}n_{\nu}$, and its trace is computed with the two dimensional
metric of the base manifold $\sigma_{\mu\nu}$ : $\mathcal{K}=\sigma^{\mu\nu
}\mathcal{K}_{\mu\nu}$. The Hamiltonian mass is then given by
\begin{equation}
M=-\frac{1}{8\pi}\lim_{r\rightarrow\infty}\int_{\mathcal{S}}\big(\mathcal{K}%
-\mathcal{K}_{0}\big)\sqrt{A(r)}\sqrt{|\sigma|}\ d\theta d\phi
\ \ ,\ \label{massdef}%
\end{equation}
where $\mathcal{K}_{0}$ is the trace of the extrinsic curvature of
$\mathcal{S}$ embedded in the background reference space-time. \newline As
pionic background, for the two solutions, the metrics (\ref{metric-bh}) and
(\ref{metric-btz}), with the vanishing mass parameters $m$ and $\mu$, are chosen.

It is worth to point that the mass can also be computed within the phase space
formalism \cite{coreani}-\cite{eloy}, giving the same result. Recently these
results have been also confirmed, thanks to counterterms methods
\cite{ortaggio}, for similar matter. \newline For the black hole solution,
using the above prescriptions we can compute the temperature, entropy and
mass. In terms of the event horizon $r_{+}$ we have, respectively
\begin{align}
T_{\text{BH}}  &  = - \frac{\Lambda r_{+}}{4 \pi} - \frac{b_{1}^{2} \kappa
K}{16 \pi r_{+}} - \frac{b_{1}^{4}\kappa K\lambda}{128 \pi r_{+}^{3}} \,
\ ,\label{TBH}\\
S_{\text{BH}}  &  = \frac{b_{2} \pi^{2} r_{+}^{2}}{2 b_{1}} \ ,\\
M_{\text{BH}}  &  = \frac{b_{2} \pi m}{4 b_{1}}\\
&  = \frac{b_{2} \pi}{384 b_{1} r_{+}} \left(  -32 \Lambda r_{+}^{4} - 24
b_{1}^{2} \kappa K r_{+}^{2} + 3b_{1}^{4} \kappa K \lambda\right)  \ ,
\label{MBH}%
\end{align}
while, for the black string with horizon radius $R_{+}$, we obtain
\begin{align}
T_{\text{BS}}  &  = -\frac{4 \Lambda R_{+}^{2} + b_{1}^{2} \kappa K}{16 \pi
R_{+}} ,\label{TBS}\\
S_{\text{BS}}  &  = \frac{\pi^{2} b_{2} \sqrt{\kappa K} R_{+}}{4
\sqrt{-\Lambda}} ,\\
M_{\text{BS}}  &  = \frac{b_{2} \pi\mu\sqrt{\kappa K} }{16 \sqrt{-\Lambda}} =
-\frac{b_{2} \sqrt{\kappa K} \pi}{64 \sqrt{-\Lambda}} \left(  2 R_{+}^{2}
\Lambda+ b_{1}^{2} \kappa K \log R_{+}\right)  . \label{MBS}%
\end{align}
In both cases these quantities satisfy the first law of black hole
thermodynamics
\begin{equation}
\delta M = T \delta S \quad.
\end{equation}

\subsection{Thermodynamical stability analysis}

From the analysis of the heat capacity, defined as
\begin{equation}
C:= T \left(  \frac{\partial S}{\partial T} \right)  \ ,
\end{equation}
we can infer the local thermodynamic stability of the two solutions. Written
it terms of the event horizon radius, $r_{+}$ and $R_{+}$ for the black hole
and the black string respectively, it reads
\begin{equation}
C_{\text{BH}} = \frac{b_{2} \pi^{2} r_{+}^{2}}{b_{1}} \left(  \frac
{32r_{+}^{4} \Lambda+ b_{1}^{2} \kappa K (8 r_{+}^{2} +b_{1}^{2} \lambda
)}{32r_{+}^{4} \Lambda- b_{1}^{2} \kappa K (8 r_{+}^{2} +3b_{1}^{2} \lambda)}
\right)  \ ,
\end{equation}
and
\begin{equation}
C_{\text{BS}} = \frac{b_{2} \sqrt{\kappa K} \pi^{2} R_{+}}{4 \sqrt{-\Lambda}}
\left(  \frac{4 R_{+}^{2} \Lambda+ b_{1}^{2} \kappa K}{4 R_{+}^{2} \Lambda-
b_{1}^{2} \kappa K} \right)  \ .
\end{equation}
The stability under thermal fluctuation occurs when the sign of the heat
capacity is positive, which means for the black hole and black string
solutions
\begin{align}
(\text{BH}) \qquad &  r_{+} > \sqrt{\frac{b_{1}^{2} \kappa K +\sqrt{b_{1}^{4}
\kappa K (\kappa K-2\lambda\Lambda)}}{-8\Lambda}} \quad,\label{rpcond1}\\
(\text{BS}) \qquad &  R_{+} > \frac{b_{1} \sqrt{\kappa K}}{2 \sqrt{-\Lambda}%
}\ . \label{rpcond}%
\end{align}
These two values coincides, when the Skyrme coupling constant $\lambda$
vanishes. Note that the above inequalities (\ref{rpcond1}) and (\ref{rpcond})
are automatically satisfied for both the black hole and black string event
horizons.\newline One might wonder if the black string might be affected by a
Gregory-Laflamme instability, where the string may collapse in a line of black
holes \cite{Gregory:1993vy}-\cite{Gregory:1994bj}.

Unfortunately, due to the complexity of the field equations, for linear
perturbations, the system can not be uncoupled to obtain a master equation for
one of the components of the perturbation; therefore it is not possible to
integrate numerically the system in the Gregory-Laflamme form and study the
unstable modes of the solutions. However, it is possible to analyze the
stability and phase transitions from the thermodynamic point of view.

\begin{figure}[h]
\includegraphics[scale=0.55]{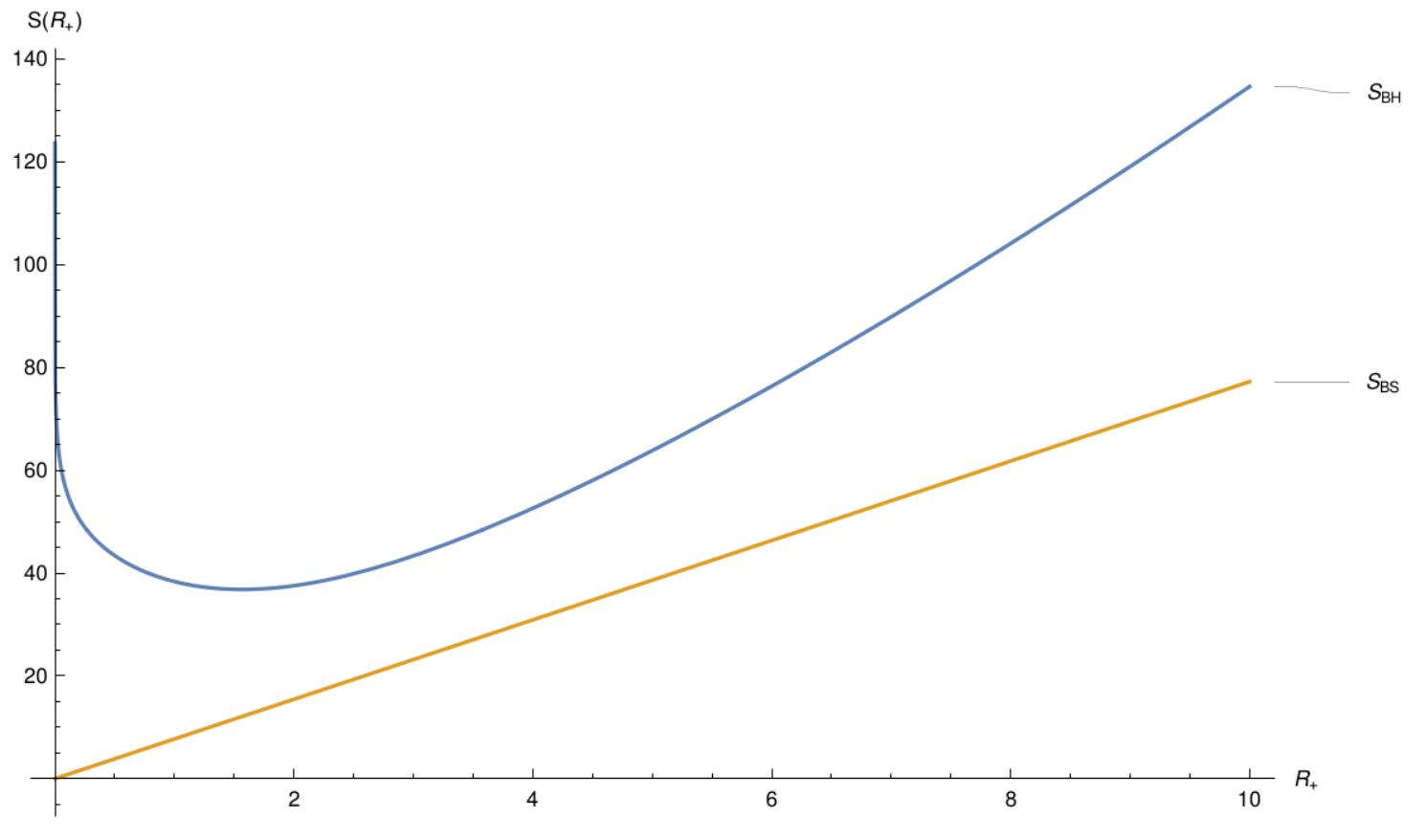}\caption{{\protect\small Entropy
of the black hole (blue line) and of the black string (yellow line) at equal
mass in terms of the string horizon $R_{+}$. The black hole entropy is always
above the string entropy, therefore, the string is not expected to decay into
the black hole configuration, for the chosen parameter set. The picture is
drawn for the fixed parameters $b_{1}=2$, $b_{2}=1$, $\Lambda=-7$, $\kappa=1$
and $K=1$, but do not change qualitatively for others admissible parametric
sets, where the entropies are well defined functions of the horizons.}}%
\label{entr-eqmass-R+}%
\end{figure}

A necessary condition for this phenomena to occur, as proposed in
\cite{gubser-mitra}, is the negativity of the quantity
\[
\frac{\partial M}{\partial S}=-\frac{b_{1}^{2}\kappa K+4R_{+}^{2}\Lambda
}{16\pi R_{+}}\quad,
\]
but in the regions of thermodynamic local stability of the string, such as the
one described by (\ref{rpcond}), it can not happen. \newline In fact, from the
inspection of the entropies of the two solutions at equal mass, we further
confirm the absence of the Gregory-Laflamme instability in our setting. More
specifically, from Eq. (\ref{MBH}) and (\ref{MBS}), we can impose the equal
mass constraint to express $r_{+}(R_{+})$ and plot both, the black hole and
black string entropy at equal mass as a function of $R_{+}$. As shown in
Figure \ref{entr-eqmass-R+}, instability is not likely to occur because the
black hole entropy, at equal mass and Skyrme parameters $b_{i}$, is always
bigger than the black string one. 

\begin{figure}[h]
\includegraphics[scale=0.44]{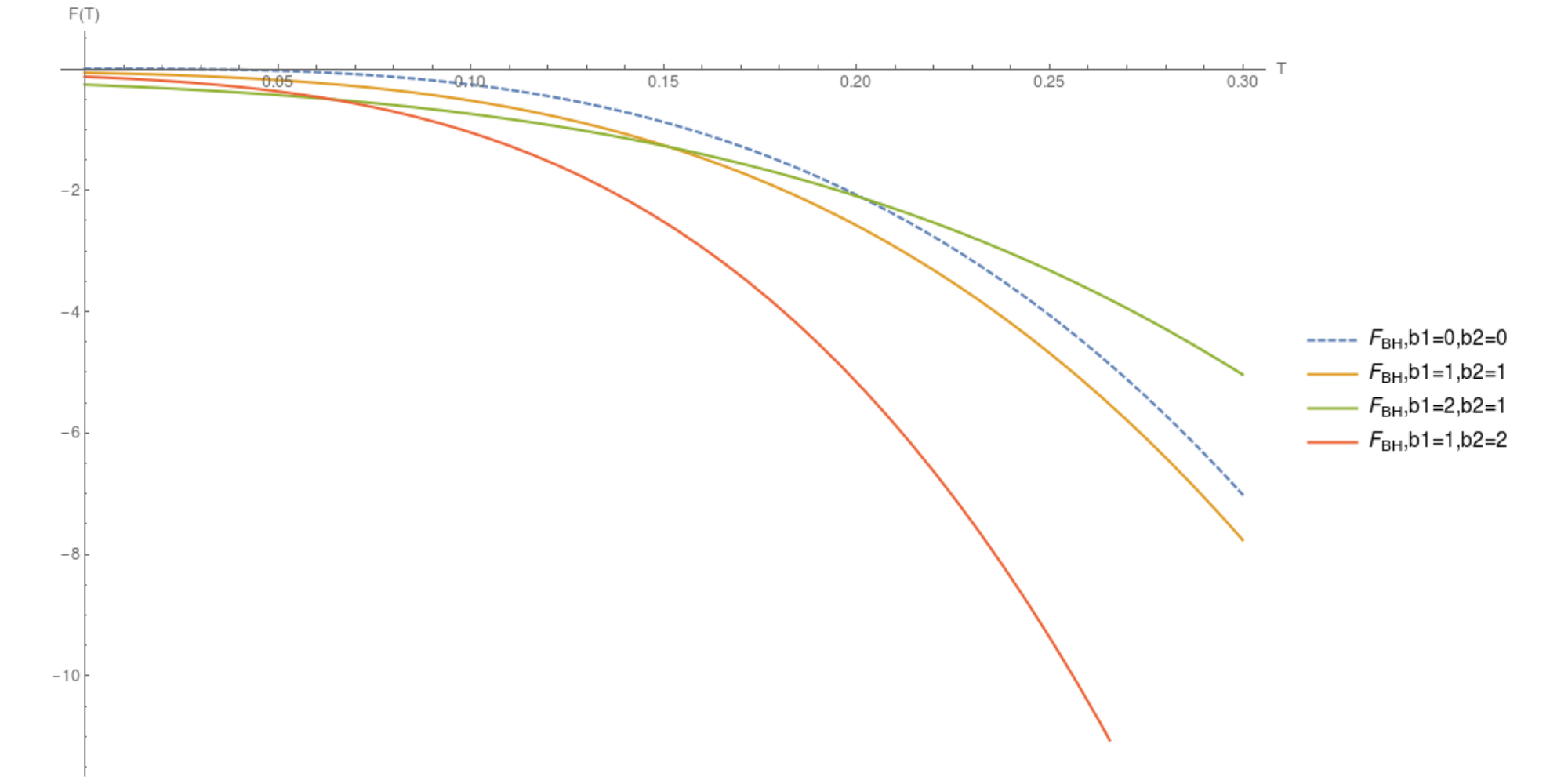}\caption{{\protect\small Free energy
$F(T)$ as a function of the temperature $T$ for the black hole configuration,
at some different values of the parameters $b_{i}$. The dashed line
corresponds to the vacuum solution which is not always favoured
thermodynamically with respect to the hairy one. In fact configurations with
$b_{1}=1$ (the red and yellow lines) have always a minor free energy with
respect to the vacuum case. Thermodynamic phase transitions can be expected,
at a certain critical temperatures located at the intersection of the
free-energy lines, for different values of the integers hairy parameters.}}%
\label{fig-fbh}%
\end{figure}

\begin{figure}[h]
\includegraphics[scale=0.44]{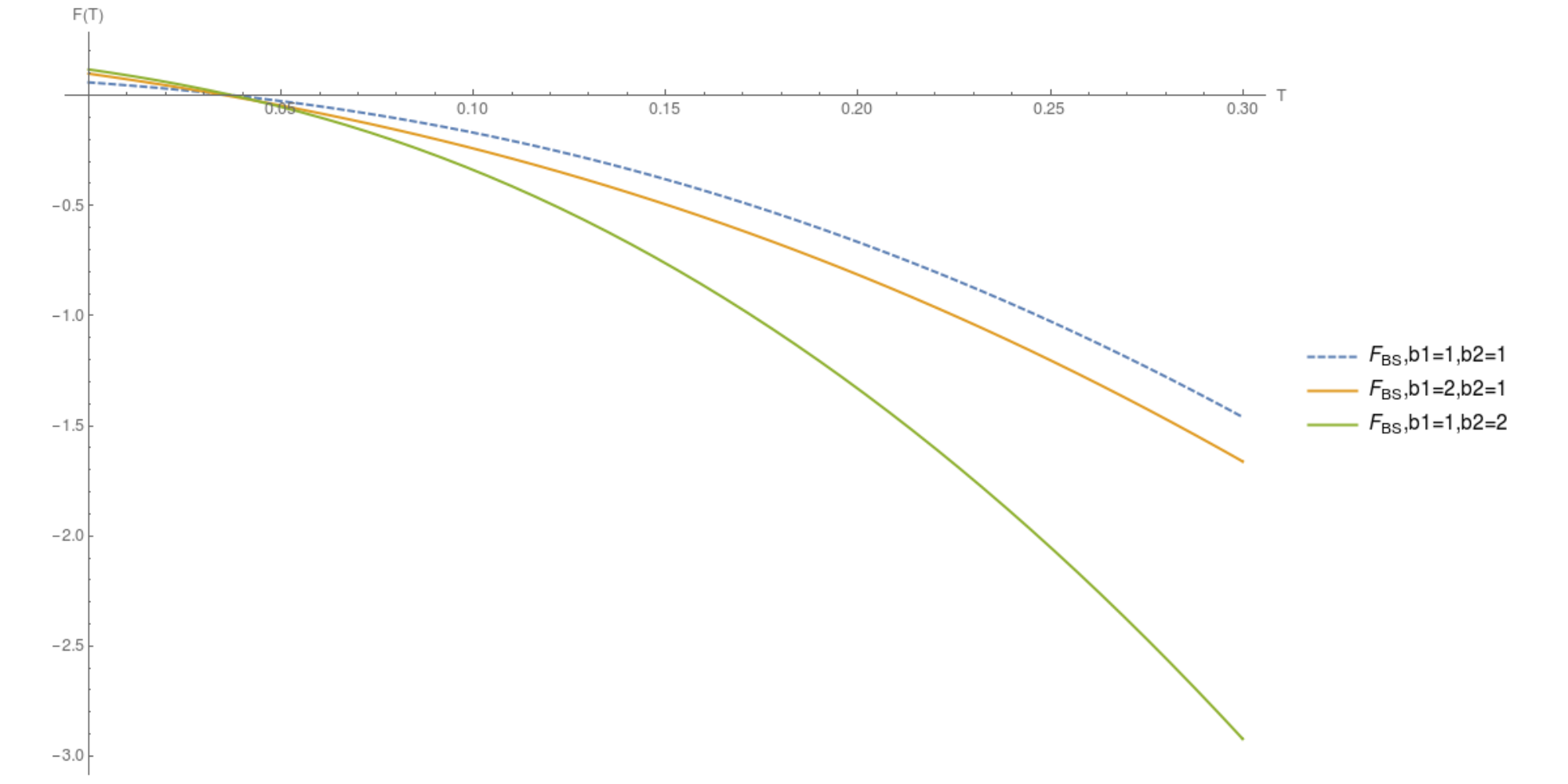}\caption{{\protect\small Free energy
$F(T)$ as a function of the temperature $T$ for the black string
configuration, at some different values of the parameters $b_{i}$.}}%
\label{fig-fbs}%
\end{figure}

Some further indication of the thermodynamic stability may come from the study
of the free energies of the two solutions. Thus, we consider the free energy
$F=M-TS$ of the black hole and of the black string, which in terms of their
event horizon are, respectively
\begin{align}
F_{\text{BH}}  &  =\frac{b_{2}\pi}{768\ b_{1}r_{+}}\left(  9b_{1}^{4}\kappa
K\lambda-24b_{1}^{2}\kappa Kr_{+}^{2}+32\Lambda r_{+}^{4}\right)
\ \ ,\label{FBH}\\
F_{\text{BS}}  &  =\frac{b_{2}\sqrt{\kappa K}\pi}{64\sqrt{-\Lambda}}%
\big[b_{1}^{2}\kappa K+2R_{+}^{2}\Lambda-b_{1}^{2}\kappa K\log(R_{+}%
)\big]\ \ . \label{FBS}%
\end{align}
To obtain the free energy in terms of the temperature is sufficient to invert
Eqs. (\ref{TBH}) and (\ref{TBS}), take the only positive root to get
$r_{+}(T)$ and $R_{+}(T$) and substitute respectively into (\ref{FBH}) and
(\ref{FBS}). The resulting analytical expression of the free energy as a
function of the temperature $F(T)$ is a little cumbersome. Thus, is more
significant to draw some picture of the free energy for some fixed values of
the parameters to appreciate, in particular, the dependence with respect to
the parameter $b_{1}$, as can be seen in Figure \ref{fig-fbh} and Figure
\ref{fig-fbs}.



We recall that $b_{1}=0$ represent the vacuum solution for the black hole
case. Configurations with bigger values of $b_{i}$ are thermodynamically
favoured (as embodied, for instance, by $b_{1}=1$ and $b_{2}=2$ in Figure
\ref{fig-fbh}) with respect to the pure gravitational solution because the
free energy is lower. The situation can change for different values of the
$b_{i}$ for the black hole case. In particular, as can be seen in the above
graphs, there are critical values of the temperature, depending on the values
of the parameters, where free energies of two different configurations
intersect, thus phase transition might be expected if discrete $b_{i}$
variations are allowed. The same qualitative behavior can be read from the
graph of the black string, but one have to remember that in this case the
comparison with the vacuum solution is not possible because proper black
string in pure general relativity without matter are not known. \newline

\section{Summary and perspectives}

In this paper we have constructed the first examples, to the best of authors
knowledge, of analytic hairy black holes with a flat toroidal horizons in the
(3+1)-dimensional Einstein $SU(2)$-Skyrme system with negative cosmological
constant. The periodic boundary conditions satisfied by the Skyrme
configurations introduce a discrete hairy parameter (as these black hole
solutions possess neither topological nor Noether charges). Such hairy
parameter can be considered neither primary (since it does not vary
continuously) nor secondary (since it can vary in a discrete set). The
solution is asymptotically locally AdS. The thermodynamics of the hairy black
hole has been analyzed in detail. The behavior one obtains is qualitatively
similar to the recent results found numerically in \cite{caldarelli} in a
different context.

Using similar techniques, we have constructed a black string in the
(3+1)-dimensional Einstein non-linear sigma model theory with negative
cosmological constant. The (2+1)-dimensional transversal sections of these
black strings correspond to a charged BTZ black hole. In this case the role of
the electromagnetic field is played by the pionic coupling constant. These
configurations can be considered as a proper black string since there is no
warping factor.

The exact hairy black hole solutions with flat horizons constructed here have
potentially applications in the context of the AdS/CFT correspondence (see
\cite{caldavero} and references therein). We hope to analyze in more details
these applications in a future publication.

Another very interesting topic is the dynamical stability analysis of the
present black holes and black strings solutions. As it has been explained in
the previous sections, this point is very difficult even from the numerical
point of view. The main issue is related to the fact that, when the Skyrme
field equations are taken into account, the fully coupled linearized
Einstein-Skyrme system cannot be reduced to a single master Schrodinger-like
equation for the perturbations as the linearized field equations do not
decouple (unlike what happens in many situations without matter fields). Thus,
one has to solve (numerically) a system of (at least) tree coupled
differential equations. We hope to come back on this issue in a future publication.

\subsection*{Acknowledgements}

{\small The authors are grateful to Julio Oliva for many enlightening comments. M.L.
and A.V. appreciates the support of CONICYT Fellowship 21141229 and 21151067,
respectively. This work has been funded by the Fondecyt grants 1160137 (FC)
and 11160945 (M.A.). M.A. is supported by Conicyt - PAI grant n$^{o}$ 79150061
. The Centro de Estudios Cient\'{\i}ficos (CECS) is funded by the Chilean
Government through the Centers of Excellence Base Financing Program of Conicyt.}

\section{Appendix}

\subsection{Main field equations}

The coupled nonlinear differential equations of the Einstein-Skyrme system for
$A(r)$, $B(r)$, $C(r)$ and $D(r)$ are given by \begin{widetext}
\begin{align}
8A^{\prime}\left( CD \right)^{\prime} +A\left[ B\left(b_2^2\kappa K(b_1^2\lambda +4C)+4D(b_1^2\kappa K+8 \Lambda C)\right)+8C^{\prime }D^{\prime}\right] &=0 \ , \\
-8BCD^2 {A^{\prime}}^2 -8ACD\left[ A^{\prime}(DB^{\prime}-BD^{\prime})-2BDA^{\prime\prime}\right] &&  \notag \\
-A^{2}\left[ B^2D(b_1^2 \kappa K (b_2^2 \lambda +4D)-4C(b_2^2 \kappa K+ 8\Lambda D)) +8CDB^{\prime}D^{\prime}+8BC({D^{\prime}}^2-2DD^{\prime\prime}) \right] &=0 \ ,  \\
-8BC^2D{A^{\prime}}^2-8ACD\left[ A^{\prime} (CB^{\prime}-BC^{\prime})-2BCA^{\prime\prime} \right]  &&  \notag \\
-A^2 \left[ B^2C(b_2^2 \kappa K (b_1^2\lambda +4C)-4D(b_1^2\kappa K+8\Lambda C)) +8CDB^{\prime}C^{\prime}+8BD({C^{\prime}}^2-2CC^{\prime\prime})\right] &=0 \ , \\
B^2CD\left[ b_1^2 \kappa K(b_2^2\lambda +4D)+4C(b_2^2\kappa K+8\Lambda D)\right] -8CDB^{\prime}(CD)^{\prime} &&  \notag \\
-8B\left[C^2{D^{\prime}}^2+D^2({C^{\prime}}^2-2CC^{\prime\prime})-CD(C^{\prime}D^{\prime}+2CD^{\prime\prime})\right] &=0 \ . \label{EE1}
\end{align}
\end{widetext}

\subsection{Gravitating regular solution}

If we choose $C(r)=1$ in the field equations and integrate the system, the
following relations
\begin{equation}
D(r)= \frac{b_{2}^{2}}{b_{1}^{2}}\ , \quad\Lambda=-\frac{1}{32}b_{1}^{2}K
\kappa(8+b_{1}^{2} \lambda)\ , \label{RS1}%
\end{equation}
reduces the system to a single equation, that can be easily solved to obtain%

\begin{equation}
A(r)=C_{1}+C_{2}r+\frac{1}{16}b_{1}^{2}K \kappa(4+b_{1}^{2}\lambda)r^{2}\ ,
\label{RS2}%
\end{equation}
with $C_{1}$, $C_{2}$ integration constants. This metric have no curvature
singularity and represents a four-dimensional space-time that is the product
of two two-dimensional space-times with constant curvature; namely
$(A)dS_{2}\times\mathbb{R}^{2}$.

When the two integration constants $C_{1}$ and $C_{2}$ are chosen
appropriately, the metric in Eqs. (\ref{metricflat1}), (\ref{RS1}) and
(\ref{RS2}) with $C(r)=1$ can be interpreted as the near horizon geometry of
the hairy black hole (analyzed in the following section).

\end{document}